\documentclass[american,english,reprint, prr, nobibnotes, anon,
superscriptaddress]{revtex4-2} 
\usepackage[T1]{fontenc}
\usepackage[utf8]{inputenc}
\usepackage{geometry}
\usepackage{babel}
\usepackage{array}
\usepackage{float}
\usepackage{booktabs}
\usepackage{multirow}
\usepackage{amsmath}
\usepackage{amssymb}
\usepackage{graphicx}
\usepackage{mathtools}
\usepackage{environ}
\usepackage{xcolor}
\usepackage[unicode=true,pdfusetitle,
 bookmarks=true,bookmarksnumbered=false,bookmarksopen=false,
 breaklinks=false,pdfborder={0 0 1},backref=false,colorlinks=false]
 {hyperref}

\makeatletter

\providecommand{\tabularnewline}{\\}

\usepackage{graphicx}
\usepackage{dcolumn}
\usepackage{bm}
\usepackage{algorithm2e}
\usepackage{algpseudocode}
\usepackage[normalem]{ulem} 
\renewcommand\vec[1]{\boldsymbol{\mathrm{#1}}}
\usepackage{todonotes}

\newif\ifshowrevisions
\showrevisionstrue
\newcommand{\hiderevisions}{\showrevisionsfalse}
\newcommand{\remove}[1]{
    \ifshowrevisions\textcolor{gray}{\sout{#1}}\else{}\fi
}
\newcommand{\add}[1]{
    \ifshowrevisions\color{blue}{#1}\color{black}\else{#1}\fi
}

\NewEnviron{removed}{
       \ifshowrevisions\textcolor{lightgray}{\BODY}\else{}\fi
}

\hiderevisions 

\makeatother

\begin{document}
\title{Temperature Steerable Flows and Boltzmann Generators}
\author{\selectlanguage{american}%
Manuel Dibak}
\thanks{These authors contributed equally to this work}
\affiliation{\selectlanguage{american}%
Department of Mathematics and Computer Science, Freie Universität Berlin, 
Arnimallee 12, 14195 Berlin, Germany}
\author{\selectlanguage{american}%
Leon Klein}
\thanks{These authors contributed equally to this work}
\affiliation{\selectlanguage{american}%
Department of Mathematics and Computer Science, Freie Universität Berlin, 
Arnimallee 12, 14195 Berlin, Germany}
\author{\selectlanguage{american}%
Andreas Krämer}
\affiliation{\selectlanguage{american}%
Department of Mathematics and Computer Science, Freie Universität Berlin, 
Arnimallee 12, 14195 Berlin, Germany}
\author{\selectlanguage{american}%
Frank Noé}
\email{frank.noe@fu-berlin.de}
\thanks{corresponding author}
\affiliation{\selectlanguage{american}%
Department of Mathematics and Computer Science, Freie Universität Berlin, 
Arnimallee 12, 14195 Berlin, Germany}
\affiliation{\selectlanguage{american}%
Department of Physics, Freie Universität Berlin, 
Arnimallee 12, 14195 Berlin, Germany}
\affiliation{\selectlanguage{american}%
Department of Chemistry, Rice University, Houston, Texas 77005, USA}
\affiliation{\selectlanguage{american}%
Microsoft Research, Station Road, Cambridge, United Kingdom}

\selectlanguage{english}%
\begin{abstract}
Boltzmann generators approach the sampling problem in many-body physics
by combining a normalizing flow and a statistical reweighting method
to generate samples in thermodynamic equilibrium. The
equilibrium distribution is usually defined by an energy function
and a thermodynamic state. Here we propose temperature-steerable flows
(TSF) which are able to generate a family of probability densities
parametrized by a choosable temperature parameter. TSFs can be embedded
in generalized ensemble sampling frameworks to sample a physical
system across multiple thermodynamic states.
\end{abstract}
\maketitle

\section{Introduction}


Sampling equilibrium states of many-body systems such as molecules,
materials or spin models is one of the grand challenges of statistical
physics. Equilibrium densities of system states $\vec{x}$ are often given
in the form
\begin{equation}
\mu_{X}(\vec{x})\propto\exp\left[-u(\vec{x})\right],\label{eq:equilibrium_density}
\end{equation}
where $u(\vec{x})$ is a reduced (unit-less) energy that combines the system's
potential $U(\vec{x})$ (if momenta are of interest we have the Hamiltonian
energy instead) with thermodynamic variables that define the statistical
ensemble. In the canonical ensemble the reduced energy is given by
$u(\vec{x})=U(\vec{x})/\tau$ where the thermal energy $\tau=k_{B}T$ is proportional
to the temperature $T$ and $k_{B}$ is the Boltzmann constant. In
order to model a system across a range of thermodynamic states, we would like to sample a
family of densities parameterized by the thermodynamic control variables
-- in the canonical ensemble,
\begin{equation}
\mu_{X}^{\tau}(\vec{x})\propto\exp\left(-\frac{U(\vec{x})}{\tau}\right).\label{eq:canonical_density}
\end{equation}
The most common approaches to sample densities (\ref{eq:equilibrium_density})
in physics and chemistry are Markov chain Monte Carlo (MCMC) or molecular
dynamics (MD) -- both proceed in steps, making small changes to $\vec{x}$
at a time, and guarantee that the target density (\ref{eq:equilibrium_density})
will be sampled asymptotically. The convergence of such sampling algorithms
is often slowed down by barriers in the energy landscape, which may
result in very long, possibly unfeasible simulation times.

Additionally, many applications require running simulations at various thermodynamic states, e.g., to study the phase behavior and temperature-dependence of materials and biological matter \cite{boettcher2018phase,wuttke2014temperature}. Moreover, generalized ensemble methods such as parallel tempering (PT) are frequently used to facilitate transitions across energy barriers and thereby enhance sampling. However, these techniques often require dozens of parallel simulations to enable Monte-Carlo exchanges between different temperatures \cite{swendsen1986replica,earl2005parallel}. 

A novel alternative to traditional sampling methods are generative machine learning models.
Recently, there has been a lot of interest in training normalizing flows
\citep{TabakVandenEijnden_CMS10_DensityEstimation,RezendeEtAl_NormalizingFlows,Papamakarios2019NormalizingFF,kobyzev2020normalizing,li2020neural}
to sample densities of many-body physics systems such as in Eq. (\ref{eq:equilibrium_density})
directly without having to run long, correlated simulation chains.
Normalizing flows \add{(see Appendix \ref{sec:flows} for a brief introduction) }transform an easy-to-sample prior distribution $p_{Z}(\vec{z})$,
e.g. a multivariate normal distribution, via a transformation $\vec{x}=f(\vec{z})$
to the output distribution $p_{X}(\vec{x})$. If $f(\vec{z})$ is invertible,
$p_{X}(\vec{x})$ can be computed by the change of variable formula
\begin{equation}
p_{X}(\vec{x})=p_{Z}(\vec{z})\left|\det J_{f}(\vec{z})\right|^{-1}.\label{eq:change_of_variable}
\end{equation}
Boltzmann Generators (BGs) \citep{noe2019boltzmann}
combine normalizing flows to minimize the distance between Eqs. (\ref{eq:equilibrium_density})
and (\ref{eq:change_of_variable}) with a statistical reweighting
or resampling method to generate unbiased samples from Eq. (\ref{eq:equilibrium_density}).
This and similar approaches have been used to sample configurations
of molecular and condensed matter systems \citep{noe2019boltzmann,wu2020stochastic},
spin models \citep{Li2018NeuralNR,Nicoli2020AsymptoticallyUE, nicoli2021estimation} and
gauge configuration in lattice quantum chromodynamics \citep{Albergo2019FlowbasedGM,Boyda2020SamplingUS, kanwar2020equivariant, albergo2021flow}. However, these previous generative approaches are only able to sample at a single predefined thermodynamic state.

This letter shows that normalizing flows can be generalized to families of ensembles across multiple temperatures and thereby greatly increase the range of thermodynamic states accessible to a sampling algorithm.
Specifically, we develop temperature-steerable
flows (TSFs) that correctly parametrize the distribution $p_{X}$ by
a temperature variable $\tau$ such that it follows Eq. (\ref{eq:canonical_density}).
We evaluate the method on the
XY model \citep{PhysRevLett.20.589} finding the correct temperature
dependence of the magnetization. 
Moreover we show for two small peptides, alanine dipeptide and tetrapeptide, that the TSF is capable of producing samples close to equilibrium at different temperatures. When trained on a single high temperature, the TSF can simultaneously sample at lower temperatures of interest, allowing a reliable estimation of physical observables and conformational distributions. Finally, due to this property, TSFs are used to facilitate exchanges in classical parallel tempering MD and thereby reduce autocorrelation times significantly.

\section{Temperature-steerable flows}

\paragraph{Temperature scaling}

Up to a normalization constant, a change from temperature $\tau$ to $\tau'$ corresponds to raising the Boltzmann distribution to the power
of $\kappa=\tau/\tau'$, $\mu_{X}^{\tau'}(\vec{x})\propto\left[\mu_{X}^{\tau}(\vec{x})\right]^{\kappa}.$
We now consider normalizing flows $f_\tau$ with priors $p_Z^\tau$ that depend on $\tau$ as a steerable parameter.
Using Eq. (\ref{eq:change_of_variable}) the output
distribution of a flow scales temperatures equally, if for any two
temperatures $\tau,\tau'$,
\begin{equation}
p_{Z}^{\tau'}(\vec{z})\left|\det J_{f_{\tau'}}(\vec{z})\right|^{-1}\propto\left[p_{Z}^{\tau}(\vec{z})\left|\det J_{f_{\tau}}(\vec{z})\right|^{-1}\right]^{\kappa}.\label{eq:scaling_condition}
\end{equation}

In this Letter we thus consider flows to be temperature steerable, if
they preserve this scaling condition. We construct flows that preserve
this proportionality in two different manners: by either keeping the
Jacobian constant and preserving the scaling condition in the prior
or selecting a constant prior and respecting the scaling condition
in the flow.

\subsection{Temperature steerable flows by volume preservation \label{subsec:Temperature-steerable-flows-by}}

\begin{figure}
\centering{}\includegraphics[width=1\columnwidth]{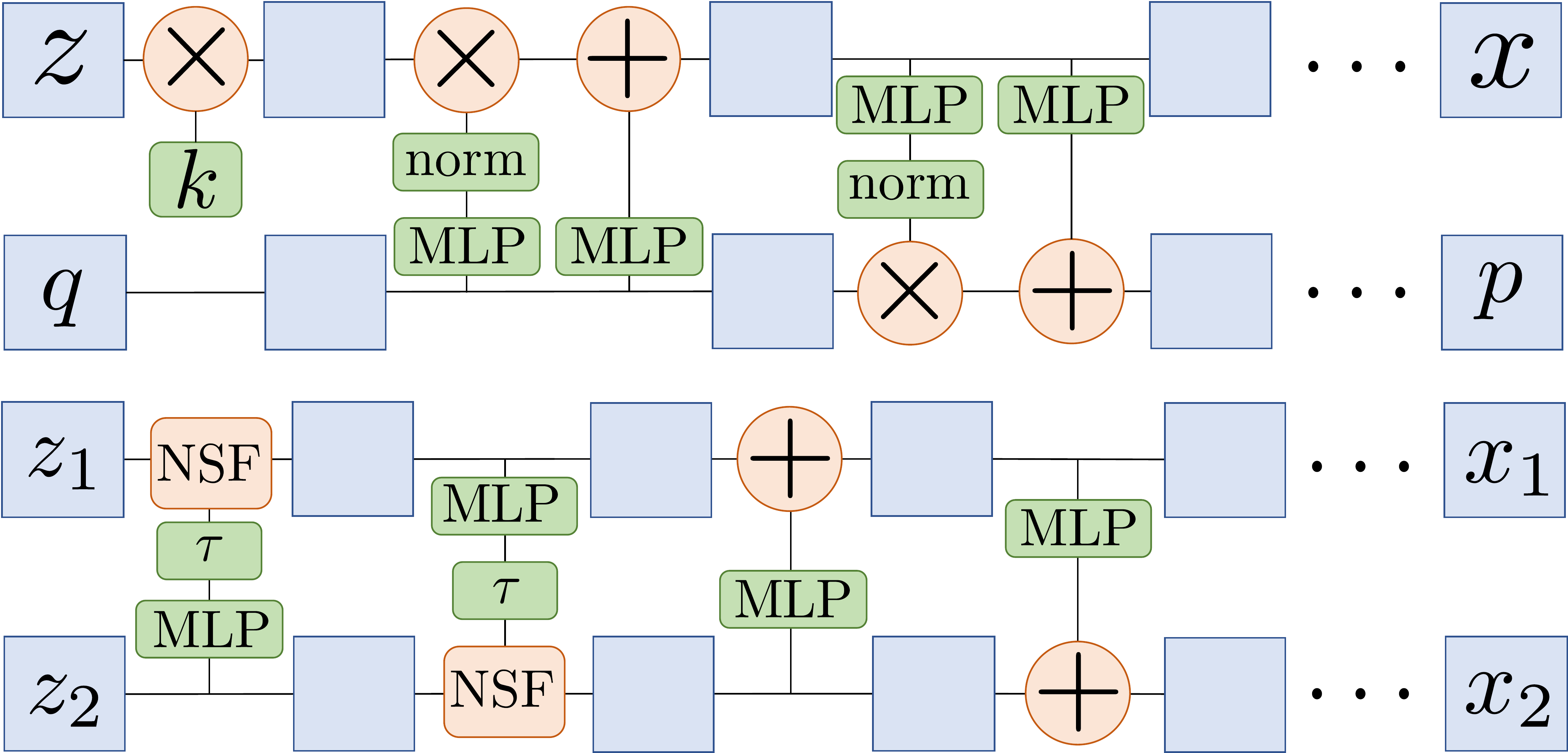}\caption{Temperature steerable flow architectures based on coupling layers, which include element-wise multiplication ($\times$) and addition ($+$). \textbf{Top:}
Auxiliary momenta $\vec{q}$ and coordinates $\vec{z}$ are coupled with volume preserving networks where the outputs of the multi-layer perceptron
(MLP) used to generate the scaling variables are normalized. The first
layer multiplies the latent space coordinates $\vec{z}$ with a scalar
factor $k$, which adjusts for the difference in entropy between latent
and phase space. \textbf{Bottom:} Temperature steerable neural spline
flows architecture. Samples from the uniform distribution are split
into two channels which are conditioned on the neural spline flows (NSF)
transformation of the other channel. The parameters for the flow are
transformed to the given temperature $\tau$. This is followed by
several layers of volume-preserving transformations \citep{dinh2016density}. \add{See Appendix \ref{sec:flows} for a description of the flow transformations. }}
\label{fig:hf_scheme}
\end{figure}

The proportionality in the prior distribution can be matched by Gaussians with variance $\tau$, i.e., $p_{Z}^{\tau}(\vec{z})=\mathcal{N}(\vec{z}\mid0,\tau)$, which
fulfills $p_{Z}^{\tau'}(\vec{z})\propto\left[p_{Z}^{\tau}(\vec{z})\right]^{\kappa}$.
This results in a condition on the Jacobian of the flow $\left|\det J_{f_{\tau}}(\vec{z})\right|^{\kappa}\propto\left|\det J_{f_{\tau'}}(\vec{z})\right|$.
Hence, flows with constant Jacobians, i.e. $\left|\det J_{f_{\tau'}}(\vec{z})\right|=\text{const.},$
are temperature steerable.


This still holds for so-called augmented normalizing flows \citep{Huang2020AugmentedNF}, where the prior and target distributions are augmented with a Gaussian distribution $p_{A}^{\tau}(\vec{q})=\mathcal{N}(\vec{q}\mid0,\tau)$ and $p_{A}^{\tau}(\vec{p})=\mathcal{N}(\vec{p}\mid0,\tau)$ respectively. This augmented flow $f_{\tau}$ is trained to match the output distribution $p_{X,A}^{\tau}(\vec{x},\vec{p})$ with the joint target distribution $\mu_{X}^{\tau}(\vec{x})p_{A}^{\tau}(\vec{p})$. 
The auxiliary variables can be interpreted as physical momenta \citep{li2020neural}, making the architecture similar to the Hamiltonian Monte Carlo method \citep{duane1987hybrid}. However, in contrast we do not propagate
the system by Hamiltonian dynamics, but learn a (deterministic) flow, as in Hamiltonian flows \citep{Greydanus2019HamiltonianNN,Toth2019HamiltonianGN}.
As we are mostly interested in the Boltzmann distribution $\mu_X(\vec{x})$ of the positions, this architecture can also be viewed as a stochastic normalizing flow \citep{wu2020stochastic}.

To generate configuration samples $\vec{x}$ from the marginal output distribution $p_{X}^{\tau}(\vec{x})$ at temperature $\tau$, we follow
three consecutive steps: (1) sample the latent configuration $\vec{z}\sim p_{Z}^{\tau}(\vec{z})$ and auxiliary momenta $\vec{q}\sim p_{A}^{\tau}(\vec{q})$, and
define the point in phase space $(\vec{z},\vec{q})$; (2) propagate the point in phase space by the flow $(\vec{x},\vec{p})=f_{\tau}(\vec{z},\vec{q})$; and (3) project onto the configuration variables $\vec{x}$

An expressive volume-preserving dynamics, i.e. $\left|\det J_{f_{\tau}}(\vec{z},\vec{q})\right|=1$,
is obtained by altering real-valued non-volume-preserving transformations
\citep{dinh2016density}, such that the product of the outputs of the scaling
layers is equal to unity. This is done by subtracting the mean of
the log outputs from each scaling layer as in Ref. \citep{Sorrenson2020DisentanglementBN}.
In addition to these volume preserving layers we scale the latent
space coordinates by a trainable scalar, which allows us to adjust for the
entropy difference between the prior and the target. The resulting
flow architecture, which still fulfills the scaling condition (\ref{eq:scaling_condition}), is shown in Fig. \ref{fig:hf_scheme} top.

As the flow fulfills
the temperature scaling condition, a temperature change of the augmented prior,
i.e., $\tau\to\tau'$, changes the output accordingly. In the case
of a factorized output distribution 
    $p_{X,A}^{\tau}\left(\vec{x},\vec{p}\right)=p_{X}^{\tau}\left(\vec{x}\right)p_{A}^{\tau}\left(\vec{p}\right)$,
the marginal output distribution $p_{X}^{\tau}\left(\vec{x}\right)$ is scaled
correctly with the temperature as well. This is ensured if the joint target distribution $\mu_{X,A}^{\tau}(\vec{x}, \vec{p}) = \mu_{X}^{\tau}(\vec{x})p_{A}^{\tau}(\vec{p})$
is matched correctly.

\subsection{Temperature steerable flows with uniform prior\label{subsec:Temperature-steerable-flows-with}}

Instead of a Gaussian prior, one can also use a uniform prior distribution on the unit box $[0,1]^{d}$ in combination with a single flow layer that scales with the temperature to construct a TSF.
While finding a flow architecture that precisely reproduces the temperature
scaling property is difficult, a good approximation is obtained using neural spline flows \citep{durkan2019neural, pmlr-v119-rezende20a}. With this type of flow
we can adjust the parameters given the temperature, such that the temperature
scaling is approximately correct (see Appendix \ref{subsec:Temperature-steerable-spline}).
In addition we combine it with volume
preserving flows, i.e. nonlinear independent
component estimation (NICE)  \citep{DinhDruegerBengio_NICE2015}, to
obtain a more expressive transformation [see Fig. \ref{fig:hf_scheme}
(bottom)].

\add{
\subsection{Training}
    As in Ref. \citep{noe2019boltzmann}, the flows are trained by a combination
    of a maximum-likelihood and energy-based loss.
    Maximum-likelihood training minimizes the negative log likelihood 
    \begin{align}
        \label{eq:nll}
        \mathcal{L}_{ML} =
        \big \langle
        & - \log p_{X,A}^{\tau}
        \big \rangle_{
            \mu_{X,A}^{\tau}
            }
        \\ 
        \nonumber
        =
        \big \langle 
            & -\log p_{Z,A}[f_{\tau}^{-1}(\vec{x},\vec{p})]
        \\ & 
            -\log \left|\det     J_{f_{\tau}^{-1}}(\vec{\vec{x},\vec{p}}) \right|
        \big \rangle_{
            \vec{x},\vec{p}\sim\mu{}_{X}^{\tau}(\vec{x})p_{A}^{\tau}(\vec{p}),
            }
        \nonumber
    \end{align}
    which agrees with the forward Kullback-Leibler divergence up to a constant.
    Computing this expectation requires samples from the product distribution $\mu{}_{X}^{\tau}(\vec{x})p_{A}^{\tau}(\vec{p})$, where the configurations $\vec{x}$ are generated by (MD) simulations, and momenta $\vec{p}$ are independent Gaussian noise. 
    
    As the target energy $u(\vec{x},\vec{p})$ is
    defined by the physical system of interest, we can also use energy-based training, which minimizes the reverse Kullback-Leibler divergence
    \begin{align}
        \label{eq:kl_loss1}
        \mathcal{L}_{KL} = 
            \big \langle
                & - \log \mu_{X,A}^{\tau} + \log p_{X,A}^{\tau}
            \big \rangle_{
                p_{X,A}^{\tau}
                }
            \\ 
            \nonumber
            = 
            \big \langle
                &
                \tau^{-1} \left(  U(\vec{x}) + \frac{\vec{|p|}^2}{2} \right)
            \\ & 
                -\log\left|\det J_{f_\tau}(\vec{z}, \vec{q})\right|
            \big \rangle_{
            \vec{z},\vec{q}\sim p_{Z}^{\tau}(\vec{z})p_{A}^{\tau}(\vec{q})
            }
            \nonumber
            \\ & + \mathrm{const.}
            \nonumber
    \end{align}
    with $(\vec{x}, \vec{p}) = f_\tau (\vec{z}, \vec{q}).$ This expectation is computed over the thermodynamic ensemble generated by the flow at a given temperature. 
    A TSF trained with Eqs. \eqref{eq:nll} and \eqref{eq:kl_loss1} will implicitly learn a representation of the Boltzmann distribution that is transferable across temperatures. It can still be useful to combine different target temperatures during training to broaden the range of temperatures at which the TSF performs well.
    
    Furthermore, we can also combine training by example and training by energy \cite{noe2019boltzmann} using a convex combination $\mathcal{L}=\left(1-\lambda\right)\mathcal{L}_{ML}+\lambda\mathcal{L}_{KL}.$
}

\subsection{Unbiased sampling: Importance sampling and latent Monte Carlo}

As in Ref. \cite{noe2019boltzmann}, we use two different methods to produce unbiased samples from the target distribution $\mu_X^\tau$. First, we employ the flow as an importance sampler and compute thermodynamic observables $\langle o \rangle_{\mu_X^\tau}$ by Zwanzig reweighting
\begin{equation}
    \langle o \rangle_{\mu_X^\tau} = \frac{
        \langle o \cdot e^{- U/\tau - \log p_X^\tau} \rangle_{p_X^\tau}
    }{
        \langle e^{-  U/\tau - \log p_X^\tau} \rangle_{p_X^\tau}
    }.
\end{equation}
Second, we extend the flow-based MCMC moves from Ref. \cite{noe2019boltzmann} to the (augmented) phase space. 

A proposal $\vec{x}'$ is generated from configuration
$\vec{x}$ by sampling auxiliary momenta $\vec{p}\sim p_{A}^{\tau}\left(\vec{p}\right),$ then applying the inverse
dynamics $(\vec{z},\vec{q})=f^{-1}_{\tau}(\vec{x},\vec{p})$, followed by a random
displacement $(\vec{z}',\vec{q}')=(\vec{z} + \Delta \vec{z} ,\vec{q}+ \Delta \vec{q})$, with $ \Delta \vec{z}, \Delta \vec{q}\sim\mathcal{N}(0,\sigma^{2})$, and finally transforming back
$(\vec{x}',\vec{p}')=f_{\tau}(\vec{z}',\vec{q}')$. Accepting such a step with probability
\begin{multline}
    p_{\text{acc}}^{\tau}\left((\vec{x},\vec{p})\to(\vec{x}',\vec{p}')\right)\\
    =\min\left\{ 1,\exp\left[-\tau^{-1}\left(U(\vec{x}')-U(\vec{x})+\frac{\left\Vert \vec{p}'\right\Vert ^{2}}{2}-\frac{\left\Vert \vec{p}\right\Vert ^{2}}{2}\right)\right]\right\} \label{eq:acceptance_probability}
\end{multline}
guarantees detailed balance in configuration space and thus ensures
convergence to the Boltzmann distribution.
As the TSF is able to generate distributions at several temperatures,
we can combine the MCMC moves with PT \citep{swendsen1986replica,Geyer1991MarkovCM,Hukushima1996ExchangeMC}.
Additionally to TSF-MCMC steps at a set of temperatures, samples can be randomly exchanged between two randomly chosen temperatures with
the usual acceptance probability. A summary of the sampling algorithm
is given in Appendix \ref{subsec:Sequential-sampling-algorithm}.

\section{Experiments}

We carry out experiments for the XY model and two small peptides, showing
that TSFs can sample the respective Boltzmann distributions at different
temperatures efficiently. The resulting flows are used to compute observables at low-temperature states from high-temperature simulations and compute temperature-dependent quantities such as absolute free energy from samples at a single thermodynamic state (see Appendix \ref{subsec:Free-energy-computation}).

\subsection{XY model}

\begin{figure}
\includegraphics[width=1\columnwidth]{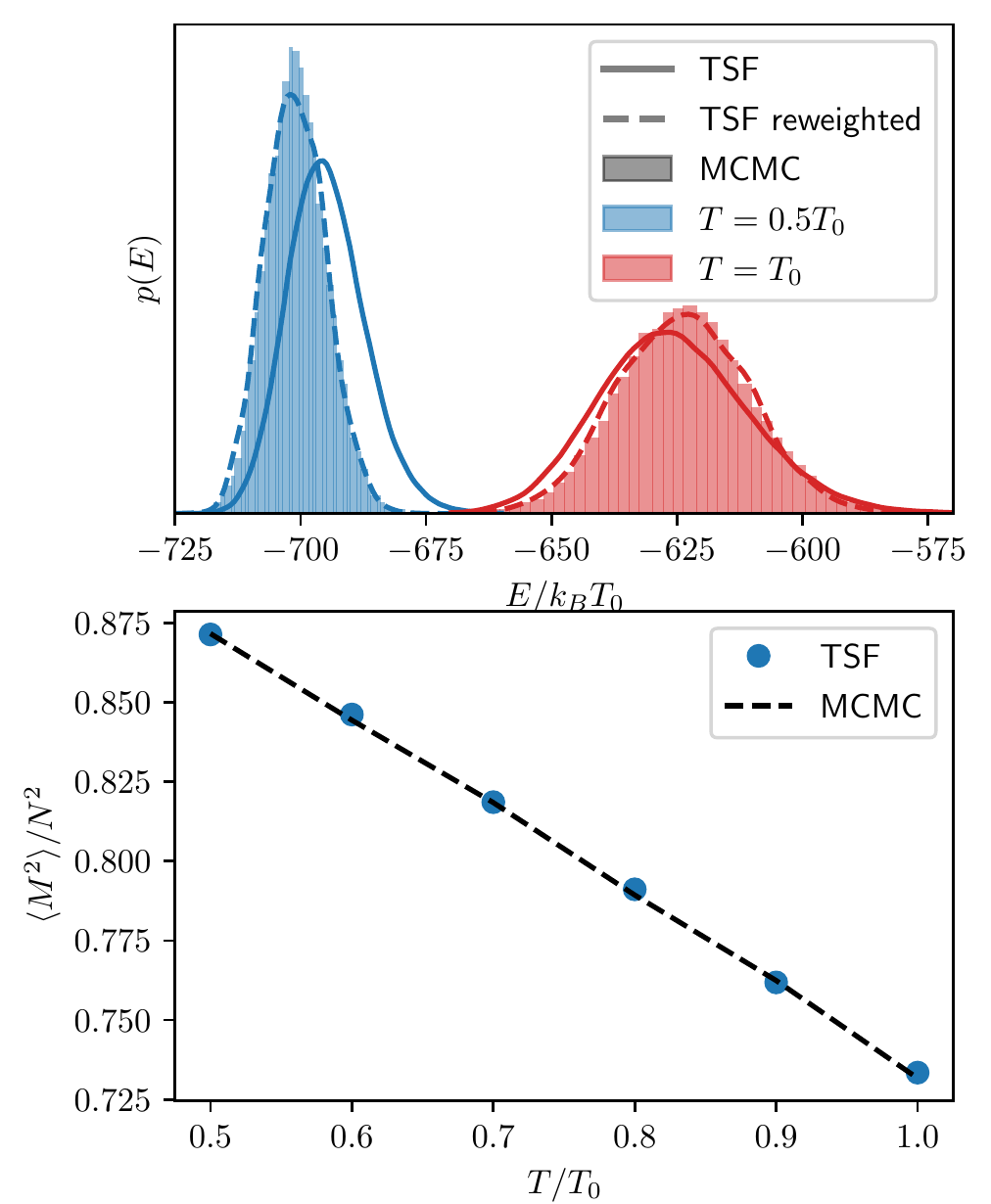}

\caption{Results of the TSF trained on the two dimensional XY model. \textbf{Top:
}Distribution of energies at the training temperatures $T_{0}$ and
\remove{the samples temperature} $0.5T_{0}$ obtained by MCMC, TSF and TSF
with reweighting. 
\textbf{Bottom: }Magnetization as a function of
the temperature compared between the TSF and MCMC samples.\label{Fig:XY_model}}
\end{figure}

As an example with angular symmetry, we investigate the XY model,
which can be considered a continuous state-space version of the Ising
model. In our experiments we consider quadratic two dimensional
lattices with $N\times N$ spins. Each spin has a continuous angle
$\theta_{i}\in[-\pi;\pi]$ and is represented by $\vec{s}_{i}=\left(\cos\theta_{i},\sin\theta_{i}\right)^{T}$.
Each spin interacts with its four nearest neighbors and an external
field $\vec{h}=\left(h,0\right).$ Hence, the Hamiltonian of the system
is given by
\begin{align*}
\mathcal{H}(s_{1},\dots,s_{N^{2}}) & =-J\sum_{<ij>}\vec{s}_{i}\cdot\vec{s}_{j}-\sum_{i}\vec{h}\cdot\vec{s}_{i}
\end{align*}
where $\sum_{<ij>}$ denotes the sum over all nearest neighbor pairs
with periodic boundaries and $J$ is the interaction constant. For
our experiments we select the parameters as $J=h=k_{B}T_{0}$ and
a lattice of $16\times16$ spins. As observable, we select the mean
squared magnetization per spin $\langle M^{2}\rangle/N^{2}=N^{-2}\left\langle \sum_{i}\vec{s}_{i}\cdot\vec{s}_{i}\right\rangle $
at a given temperature. For producing reference configurations,
we use long runs of Glauber dynamics \citep{glauber1963time}. The
TSF consists of a uniform prior, a temperature scaled NSF, followed
by seven blocks of circular NICE [see Fig. \ref{fig:hf_scheme} (bottom) and Appendix \ref{subsec:Detailed-description-of} for details].
Training is performed solely with the energy-based loss [Eq. \eqref{eq:kl_loss1}]. \add{Since NSFs are only approximately temperature scaling, }we use a convex combination of temperatures $T=\{0.5,1.0,1.3\}T_{0}$ for training.
We generate data sets at temperatures ranging $\text{0.5\,\ensuremath{T_{0}},0.6\,\ensuremath{T_{0}},\ensuremath{\dots},1.0\,\ensuremath{T_{0}}}$
and observe an excellent overlap of the reweighted energies at the highest
and lowest temperature with the reference configurations [Fig. \ref{Fig:XY_model}
(top)]. Furthermore, we compare the mean squared magnetization per spin
and again find excellent agreement between TSF and Glauber dynamics
[Fig. \ref{Fig:XY_model} (bottom)].

\subsection{Alanine di- and tetrapeptide}
\begin{figure*}
\begin{centering}
\includegraphics[width=0.89\textwidth]{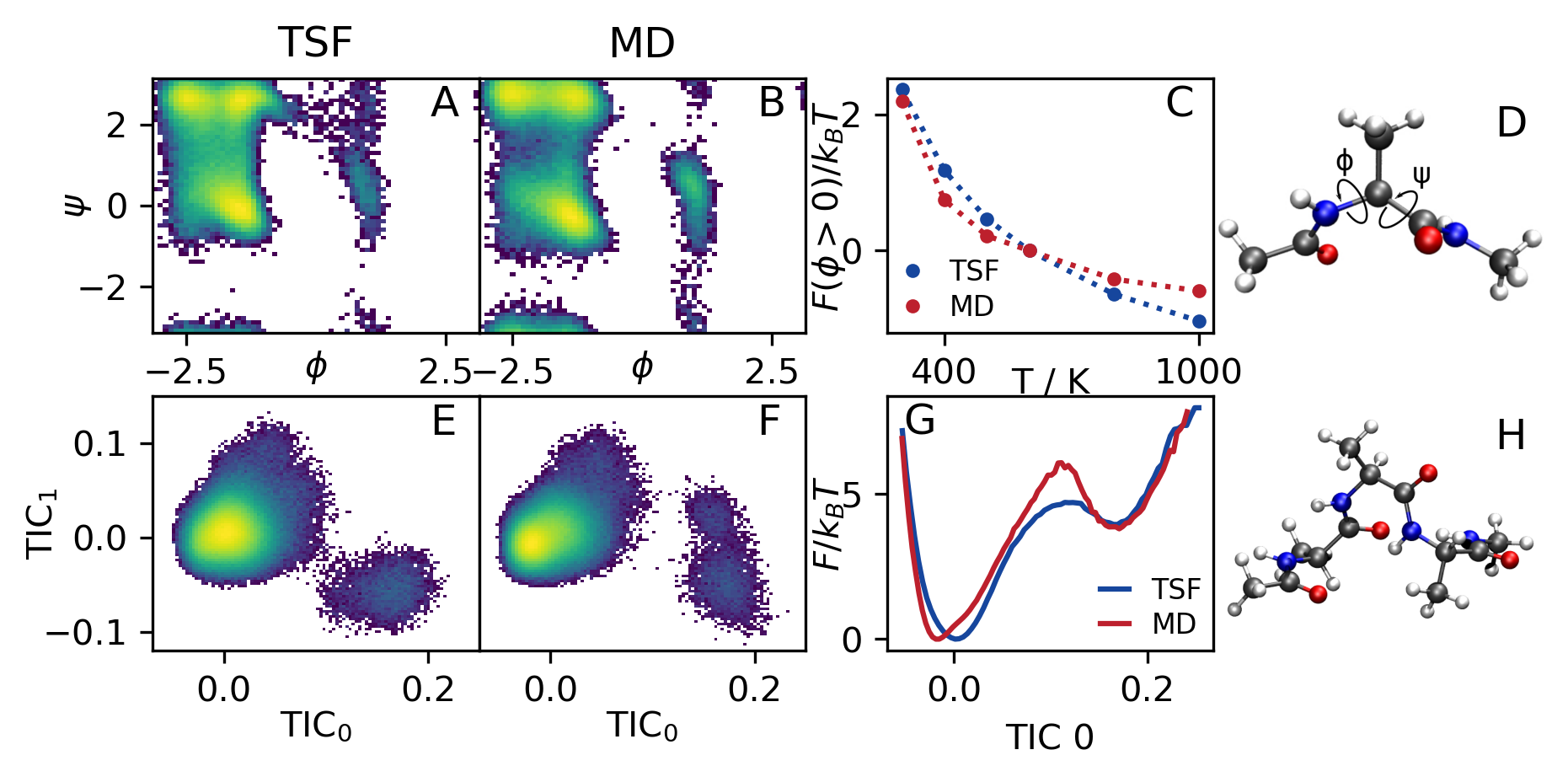}
\par\end{centering}
\caption{Results for alanine dipeptide (ala2) (\textbf{D}, first row) and alanine tetrapeptide (ala4) (\textbf{H}, second row) in implicit solvent. All TSFs are trained from samples at 600K \textbf{A,
B:} Ramachandran plots produced by the TSF and MD at 300K. \textbf{C:} Comparison of the free energy difference of the two metastable states along the $\phi$ axis.
\textbf{E, F:} Comparison of the first two  time-independent components (TICs) of ala4 at the sampling temperature of 300K. Figure \textbf{G} The free energy along the first TIC.
\label{fig:ala2}}
\end{figure*}
\begin{figure}
\includegraphics[width=1\columnwidth]{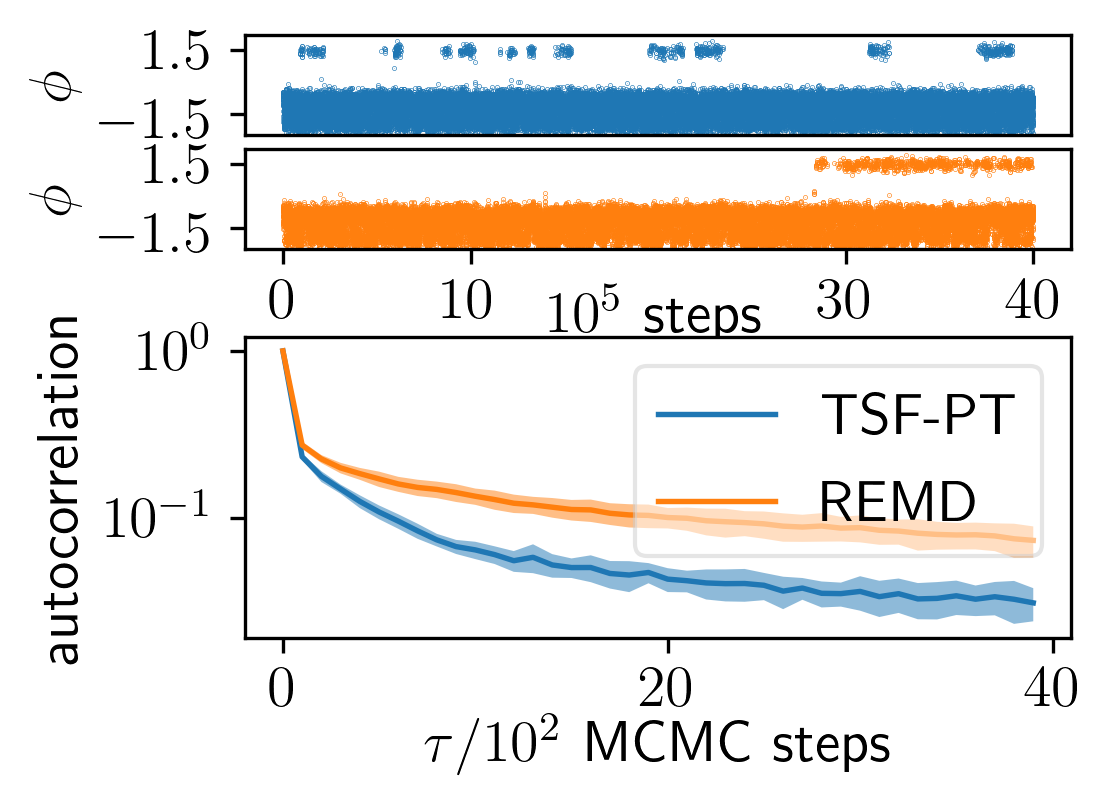}

\caption{Comparison of replica exchange molecular dynamics simulation (REMD)
and the TSF with parallel tempering (TSF-PT), operating in a PT scheme
on $8$ different temperatures in the range between $300$
and $600\,\mathrm{K}$\textbf{ Top: }One example trajectory of $\phi$
angles of the $10$ independent runs. Despite being over two times
longer, no transitions are observed in three out of the ten REMD runs,
while all of the TSF-PT trajectories cross many times between the
two metastable states. \textbf{Bottom:} Autocorrelation of the $\phi$
angle as a function of underlying MCMC/MD steps. The autocorrelation
function decays more rapidly in the TSF-PT method. This hints toward
this method being more sampling efficient. Non traversing trajectories
of the REMD method were excluded in the calculation of the autocorrelation.\label{fig:Top:-Summarized-trajectories}}
\end{figure}

We further test TSFs on the alanine di- and tetrapeptide molecules in an implicit
solvent model. For this system we use an invertible coordinate transformation
layer and operate the TSF on a representation of the molecule in terms
of distances and angles. Our goal is to use samples at $T=600\:\mathrm{K}$
to train the TSF and then use the TSF to sample at $T=300\:\mathrm{K}$,
comparing it to a MD simulation at $T=300\:\mathrm{K}$. 

\paragraph{Alanine tetrapeptide}
With the TSF we are able to generate samples at $T=300\:\mathrm{K}$ that closely resemble the equilibrium distribution. To demonstrate this, we project the configurations into the space of the slowest transition between states. These are determined by a time-independent component analysis (TICA \cite{perez2013identification}) from an exhaustive MD simulation at $T=300\:\mathrm{K}$. We observe generally good agreement with MD simulations at the low temperature (Figs. \ref{fig:ala2} E-G) in the relevant slowest coordinates, while slightly underestimating the barrier height (Fig. \ref{fig:ala2} G).

\paragraph{Alanine dipeptide}
We use the TSF to generate samples in configuration space and compare the Ramachandran plots. At $T=300\:\mathrm{K}$ (Figs. \ref{fig:ala2} A and B) the TSF still finds the major minima at around $\phi\approx-2$,
but under-samples the minimum at $\phi\approx1$. This deviation from
the target distribution likely stems from limited expressivity
of the flow.
We further utilize the TSF to compare the free energy difference of the two states along the $\phi$ axis as a function of the temperature (Fig. \ref{fig:ala2} C). We observe an exact match at the training temperature and slight deviations when moving away from it.
To recover the the correct distribution along the $\phi$ angles,
we use the Monte Carlo scheme in a PT fashion with eight temperatures
in the range $300$-$600\,\mathrm{K}$ [see Fig. \ref{fig:alanine_details} (bottom)].

To assess the efficiency of the sequential sampling procedure, we
compare it to the replica exchange molecular dynamics simulation (REMD)
at the same temperatures. We observe that for ten independent runs,
in REMD only seven transition between the metastable states within $10\,\mathrm{M}$
steps, while all $10$ transition with the TSF-PT method, which additionally
only consists of $4\,\mathrm{M}$ steps [Fig. \ref{fig:Top:-Summarized-trajectories}
(top)]. Furthermore, the autocorrelation of the slowest process [Fig.
\ref{fig:Top:-Summarized-trajectories} (bottom)], which are the transitions
along the $\phi$ angle, decays considerably faster in the TSF-PT
method. In addition, we compare the methods based on their sampling efficiency $\eta=N_{\mathrm{eff}}/N$,
where $N$ is the number of underlying MCMC steps and $N_{\mathrm{eff}}$
is the effective sample size (see Appendix \ref{subsec:Autocorrelation-analysis-of}
for details). Table \ref{tab:efficiency}
shows that the TSF-PT method produces independent samples at about
four times the rate of the REMD method.

\begin{table}
\caption{Efficiency as the number of effective steps per underlying sampling
step for different sampling methods.}
\begin{tabular}{ccccc}
\toprule 
Method: & TSF-PT & REMD & MD(600K) & MD(300K)\tabularnewline
\midrule
$\eta\times10^{4}$ & 1.36 & 0.32 & 0.38 & 0.02\tabularnewline
\bottomrule
\end{tabular}\label{tab:efficiency}
\end{table}

\section{Discussion}

In this Letter, we derived and constructed temperature steerable flows
(TSFs) that correctly scale the output distribution of a BG with temperature.
To this end we formulated a condition for such flows and introduced two different methods of constructing them. We showed that this type of flow can
be used to train a BG at one temperature and generate distributions
at lower temperatures.

For the XY model we were able to predict the
correct temperature dependence of the magnetization. Furthermore,
we showed for alanine dipeptide that the efficiency of parallel tempering
can be improved by using our TSF for the MCMC proposals at different
temperatures. Further progress could be made by combining samples
at different temperatures when collecting training data and thus improve
the quality of the BG.

\add{
While the presented results demonstrate the promise and uniqueness of the TSF method, practical applications to high-dimensional physical systems of interest will likely require further modifications to the network architecture.
Future work should consider combining TSF with conditioner networks that respect the symmetries of the potential energy.
In this spirit, }the presented temperature-scaling property complements existing equivariant flows that maintain group transformations
such as rotation and permutation \citep{Khler2020EquivariantFE,Rezende2019EquivariantHF,zhang2018monge,satorras2021n}.

\begin{acknowledgments}
\section{Acknowledgment}
We gratefully acknowledge support by the Deutsche Forschungsgemeinschaft
(SFB1114, Projects No. C03, No. A04, and No. B08), the European Research Council (ERC CoG 772230
\textquotedblleft ScaleCell\textquotedblright ), the Berlin Mathematics
center MATH+ (AA1-6), and the German Ministry for Education and Research
(BIFOLD - Berlin Institute for the Foundations of Learning and Data).
We thank Jonas Köhler, Michele Invernizzi, and Yaoyi Chen for insightful
discussions. Moreover, we thank the anonymous reviewers for their constructive feedback. 
\end{acknowledgments}

\newpage{}

\bibliographystyle{unsrt}
\bibliography{literature}

\clearpage{}

\appendix

\add{
    \section{Normalizing flows (NICE and NSF)}
    \label{sec:flows}
    Normalizing flows \cite{TabakVandenEijnden_CMS10_DensityEstimation,tabak2013family,Papamakarios2019NormalizingFF} are invertible neural networks $f$ that operate as density maps on top of a prior distribution $p_X$. They are usually designed such that the following numerical operations are computationally efficient:
    \begin{itemize}
        \item forward evaluation $\vec{x} = f(\vec{z})$
        \item inverse evaluation $\vec{z} = f^{-1}(\vec{x}),$
        \item evaluation of the Jacobian determinant $|\det J_f(\vec{z})|$  and, by extension, its inverse $|\det J_{f^{-1}}(\vec{x})|,$ and
        \item computation of the following parameter gradients for training (``backward evaluation'')
        \begin{equation*}
            \nabla_{\vec{\theta}} f,\ \nabla_{\vec{\theta}} f^{-1},\ 
            \nabla_{\vec{\theta}} |\det J_f|,\  \nabla_{\vec{\theta}} |\det J_{f^{-1}}|.
        \end{equation*}
    \end{itemize}
    An important network architecture that meets these requirement are coupling layers, which are algorithmically similar to reversible intergrators. They operate on two input vectors (e.g. positions $\vec{z}$ and momenta $\vec{q}$), where only one input (e.g. positions) is transformed by an element-wise transform $\vec{x} = s(\vec{z}; \vec{\vartheta}).$ The transform $s$ is a function that has cheap derivatives and inverse. NICE \cite{dinh2016density} corresponds to $s$ being a simple sum $s(\vec{z}; \vec{\vartheta}) = \vec{z} + \vec{\vartheta}.$ Neural spline flows (NSF) \cite{durkan2019neural} correspond to $s$ being strictly increasing rational quadratic splines, where $\vec{\vartheta}$ contains the spline knots and slopes (see below).
    
    Crucially, the transform parameters $\vec{\vartheta}$ are generated by a separate neural network, $\vec{\vartheta} = c(\vec{q}; \vec{\theta}).$ This conditioner network $c$ need not be invertible and can be a simple multilayer perceptron. The term coupling layer denotes the overall trainable transformation 
    \begin{align*}
            \left(
            \begin{array}{c}
                 \vec{z}   \\
                 \vec{q}
            \end{array}
            \right)
        \mapsto
            \left(
            \begin{array}{c}
                 \vec{x}   \\
                 \vec{p}
            \end{array}
            \right)
        = 
                \left(
            \begin{array}{c}
                 s(\vec{z}; c(\vec{q}; \vec{\theta}))   \\
                 \vec{q}
            \end{array}
            \right),
    \end{align*}
    which allows efficient inversion and computation of the Jacobian determinant. Stacking such coupling layers and reversing the roles of $\vec{q}$/$\vec{p}$ and $\vec{z}$/$\vec{x}$ in between yields invertible neural networks that can express complicated diffeomorphisms.

}

\section{Temperature steerable spline flows\label{subsec:Temperature-steerable-spline}}

\begin{figure}[b]
\includegraphics[width=1\columnwidth]{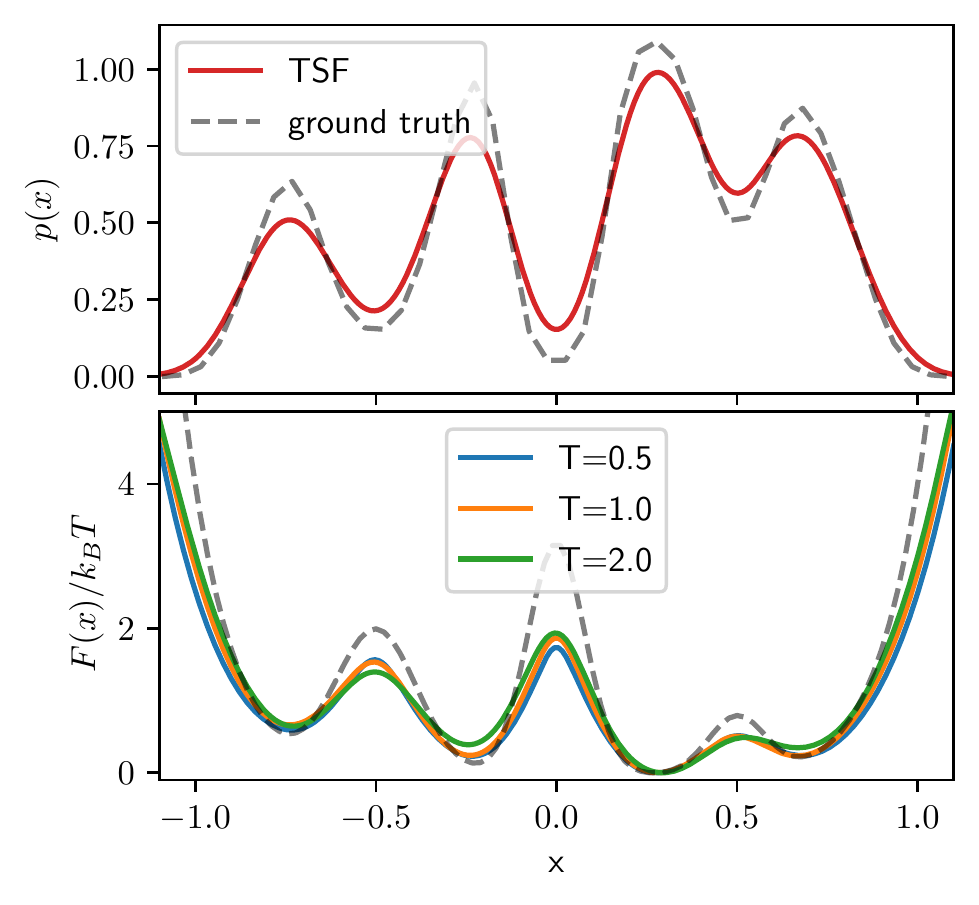}\caption{Temperature steerable spline flow as a trained inverse sampler for
1D densities. The free energies are given in units of $k_{B}T$. In
these units the ground truth coincides for different temperatures.
The top two figures show an application to the asymmetric double well
potential and the bottom two to the Prinz potential \citep{prinz2011markov}.\label{fig:Temperature-steerable-spline}}
\end{figure}

In a spline flow transformation, each element of the input vector
is transformed via an invertible scalar function $y_{i}=F(x_{i})$
defined on the unit interval $F:[0,1]\to[0,1]$, s.t. $F(0)=0$ and
$F(1)=1$. Neural Spline Flows are especially useful when transforming a quantity with circular symmetry as they can easily be adjusted to satisfy the periodicity of the variables. For this transformation to be invertible, it needs to be monotonous. It can be interpreted as the cumulative distribution function
corresponding of a probability density function $p$ defined on the
unit interval. The function $F(x)$ is approximated by a spline $s(x)$,
a piecewise defined function, which is invertible in each interval.
The spline is parametrized by $K$ coordinates, and slopes $(x^{(i)},y^{(i)},\delta^{(i)})$
which are the function values and derivatives of $y^{(i)}=F(x^{(i)})$,
$\delta^{(i)}=p(x^{(i)})$ . The spline then interpolates the function
values in between these coordinates. There exists a whole range of
different ways to define a spline. The recently proposed Neural Spline
Flows (NSF) \citep{durkan2019neural} use quadratic rational splines
and have been shown to perform best in a series of tasks. Thus these
have been utilized in this work.

A quadratic rational spline can alternatively be defined by $N$ bin
widths, heights and slopes $(w_{i,}h_{i},\delta_{i})$, with $\sum_{i}w_{i}=1$,
$\sum_{i}h_{i}=1$. Assuming that the spline is an approximation of
$F(x)$, we find that $h_{i}=\int_{x_{i}}^{x_{i}+w_{i}}p(x)dx$, with
$x_{i}=\sum_{j=1}^{i-1}w_{i}$. For the temperature scaling to hold,
we assume that $p(x)$ is of the form $p(x)=\exp(\beta v(x))$ for
some continuous function $v(x)$. We then see that $\delta_{i}=\exp(\beta v_{i})$,
with $v_{i}=v(x_{i})$ and can approximate by the mean value theorem
$h_{i}=\int_{x_{i}}^{x_{i}+w_{i}}p(x)dx\approx\exp(\beta\tilde{v_{i}})w_{i}$,
with $\tilde{v}_{i}=v(\xi)$ for $\xi\in[x_{i},x_{i+1}]$. With these
assumptions it is clear how scale the spline parameters to other temperatures
\begin{align*}
h_{i}^{\tau} & =\exp(\tau\tilde{v}_{i})w_{i}/\sum_{i}\exp(\tau\tilde{v}_{i})w_{i},\\
\delta_{i}^{\tau} & =\exp(\tau v_{i}).
\end{align*}
Thus the transformation is fully parametrized by the set of values
$(w_{i},v_{i},\tilde{v_{i}})$. To demonstrate that this indeed produces
a temperature steering transformation of the variable $\vec{x}$, we apply
this method to a test system, namely the Prinz potential \citep{prinz2011markov}.
Here we only consider one transformation with fixed weights. We train
the set of parameters directly by the KL-loss. We generally observe
a good fit to the ground truth (Fig. \ref{fig:Temperature-steerable-spline}).
The temperature steering property is evident by observing, that the
free energy almost coincides at the three different temperatures,
when expressed in terms of thermal energy.

For higher dimensional systems one makes use of coupling layers. For
the temperature scaling condition to hold, we can only use one coupling
layer on each subset of coordinates. To further enable the transformation
to capture correlations in the system, we combine the procedure with
volume preserving transformations.


\section{Sequential sampling algorithm\label{subsec:Sequential-sampling-algorithm}}

\begin{algorithm}
\label{alg:sampling}
\DontPrintSemicolon
\caption{Sampling algorithm used in the parallel tempering Ala2 system.}
\SetKwInOut{Input}{input}
\SetKwInOut{Output}{output}
\Input{$l_{s}=[~]: \text{empty list for samples}$ \\
$\tau : \text{list of }N_T\text{ temperatures}$\\
$\mathrm{state} \gets \textrm{init state}: N_{\tau} \text{ initial configurations}$\\
$N_{\textrm{iterations}}: \text{number of generated samples}$\\
$n_{\textrm{propagate}}: \text{number of propagation steps}$\\
$n_{\textrm{swap}}: \text{number of temperature swaps}$
}
\For{$i \gets 1$ to $N_{\textrm{iterations}}$}{
	\For{$j \gets 1$ to $n_{\textrm{propagate}}$}{
		\For{$k \gets 1$ to $N_{\tau}$}{
			$\vec x \gets \textrm{state}_k$\;
			$\vec{p} \gets \text{sample from } p_{A}^{\tau_{k}} $\;
			$(\vec z, \vec q) \gets D^{-1}((\vec x,\vec p))$\;
			$\vec \omega \gets \text{sample from } \mathcal{N}(0, \sigma \mathbb{1})$\;
			$(\vec z \, ', \vec q \, ') \gets (\vec z, \vec q) + \vec \omega$\;
			$(\vec x \, ',\vec p \, ') \gets D((\vec{z}\, ', \vec{q}\, '))$\;
			$p_{\text{acc}} \gets p_{\mathrm{acc}}^{\tau_k}((\vec x, \vec p) \to (\vec x \, ', \vec p \, '))$ (Eq.
\ref{eq:acceptance_probability})\;
		\If{$r\sim \mathcal{U} (0,1) < p_{\text{acc}}$}{ $\text{state}_k \gets \vec{x}\, '$ }
	}
}
	\While{$j<n_{\textrm{swap}}$}
	{
		randomly select $\alpha, \beta \leq N_{\tau}$, $\alpha \neq \beta$\;
		$\vec x \gets \textrm{state}_{\alpha}$\;
		$\vec y \gets \textrm{state}_{\beta}$\;
		$p_{\text{acc}} \gets p_{\mathrm{acc}}(\vec{x}, \vec{y}, \tau_i, \tau_j)$ (Eq. \ref{eq:temperature_swap})\;
		\If{$r\sim \mathcal{U} (0,1) < p_{\text{acc}}$}{ 
		$\text{state}_\alpha \gets \vec{y}$\;
		$\text{state}_\beta \gets \vec{x}$\;
		}
	}
	$l_{\text{s}}\text{.append }(\vec{x})$\;
	$i \gets i+1$\;
}
\Output{list of samples $l_{\text{s}}$}
\label{sampling_algorithm}
\end{algorithm}The TSF-PT algorithm is shown in Algorithm \ref{sampling_algorithm}. For
a random swap of temperatures between two samples $\vec{x}_{i}$ and
$\vec{x}_{j}$ at temperatures $\tau_{i}$ and $\tau_{j}$, detailed
balance is preserved by the acceptance probability

\begin{equation}
p_{\mathrm{acc}}=\min\left\{ 1,\exp\left[\left(U(\vec{x}_{i})-U(\vec{x}_{j})\right)\left(\frac{1}{\tau_{i}}-\frac{1}{\tau_{j}}\right)\right]\right\} .\label{eq:temperature_swap}
\end{equation}

\section{Autocorrelation analysis of Ala 2 runs\label{subsec:Autocorrelation-analysis-of}}

\begin{figure*}
\includegraphics[width=0.45\textwidth]{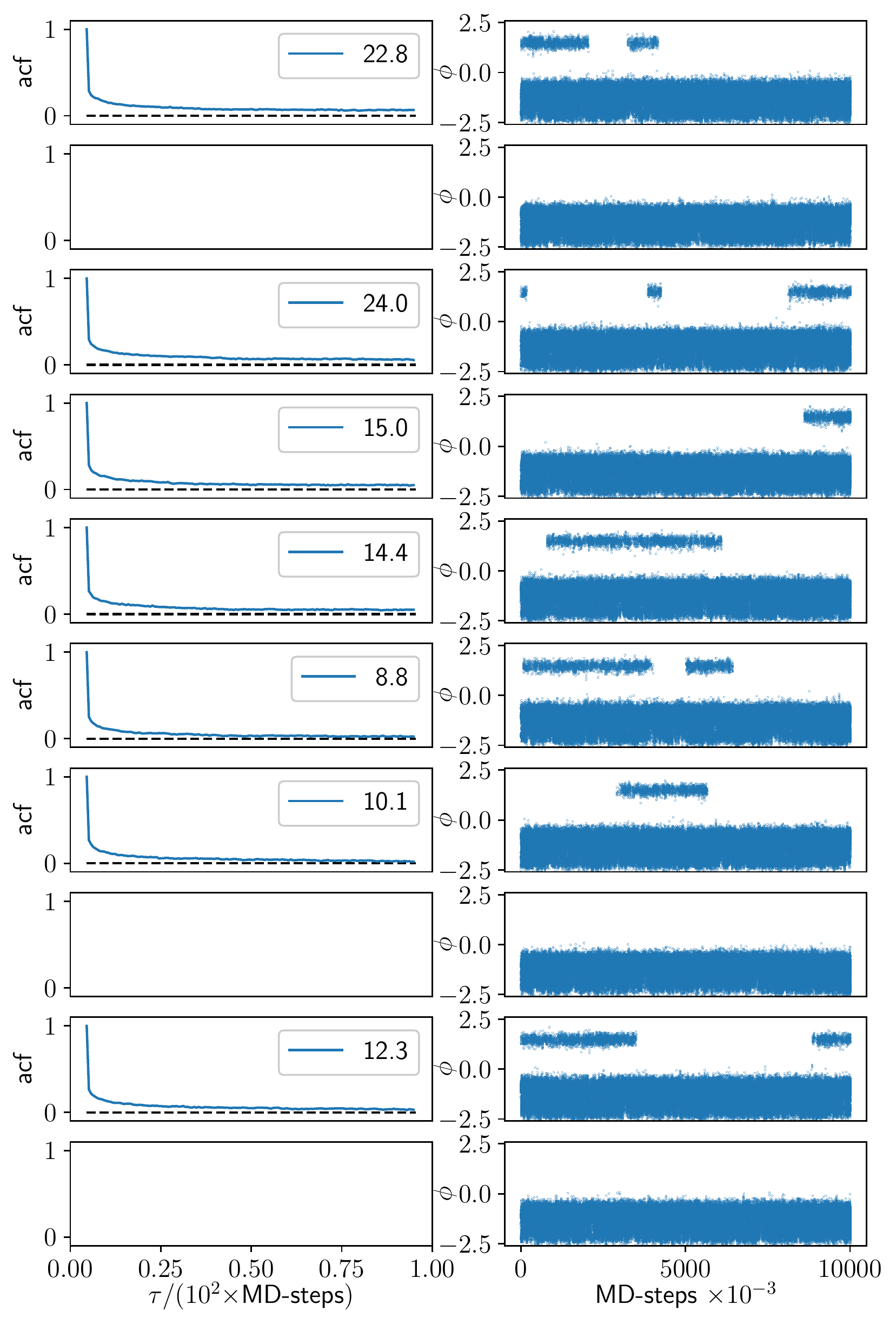}\includegraphics[width=0.45\textwidth]{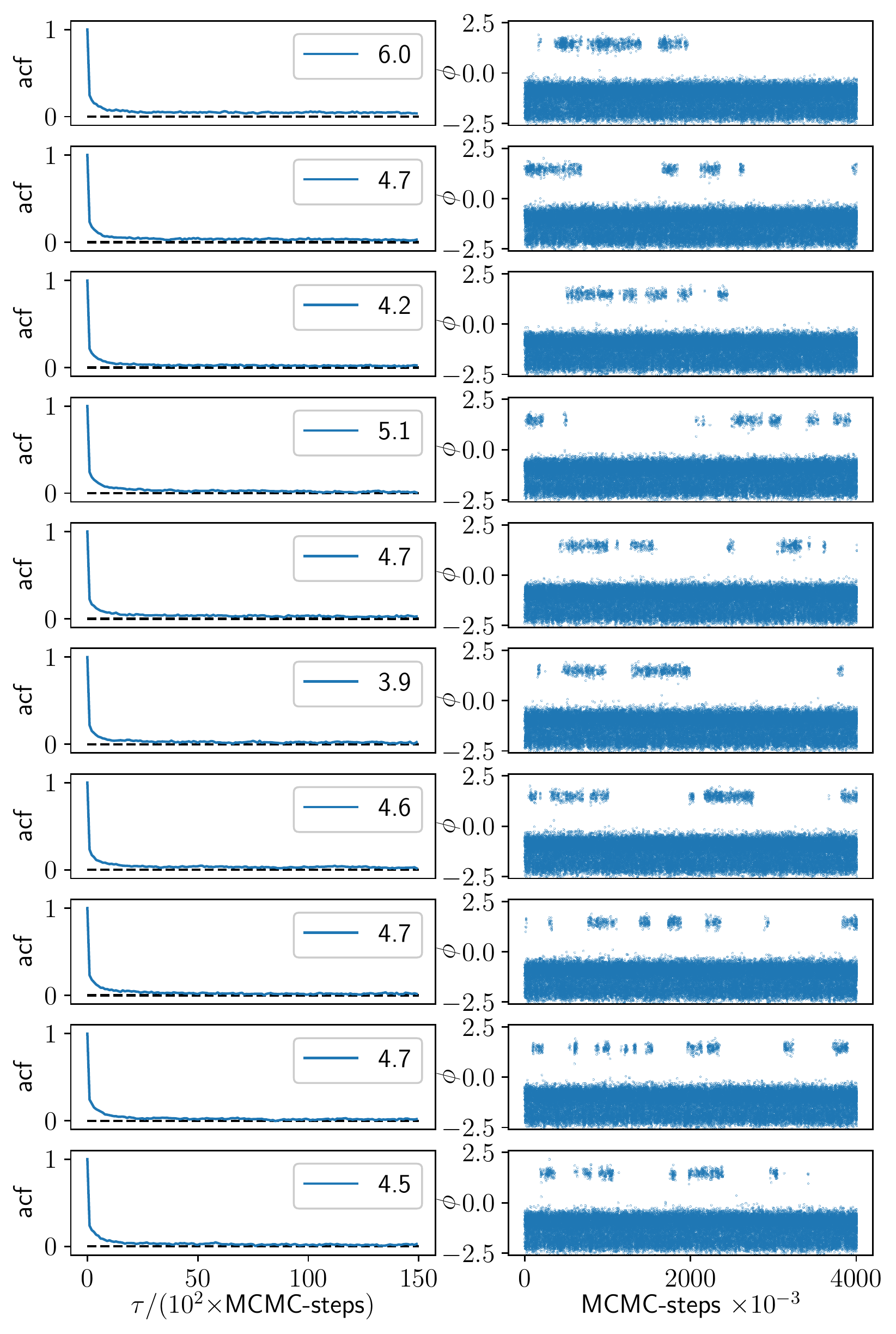}\caption{Trace plots and autocorrelation of the $\phi$ angle in the sampling
runs of the Ala2 molecule at $T=300\,\mathrm{K}$. Left for replica
exchange molecular dynamics (REMD), right for TSF-PT\label{fig:Trace-plots-and}.
The numbers in the legends of the autocorrelation function (acf) plots
are the integrated autocorrelation times $\tau$. We observe more
frequent transitions between the metastable states at $\phi<0$ and
$\phi>0$ in the TSF-PT method. This is also reflected in the integrated
autocorrelation times, which are considerably lower for TSF-PT.}
\end{figure*}

\begin{figure}
    \includegraphics[width=0.8\columnwidth]{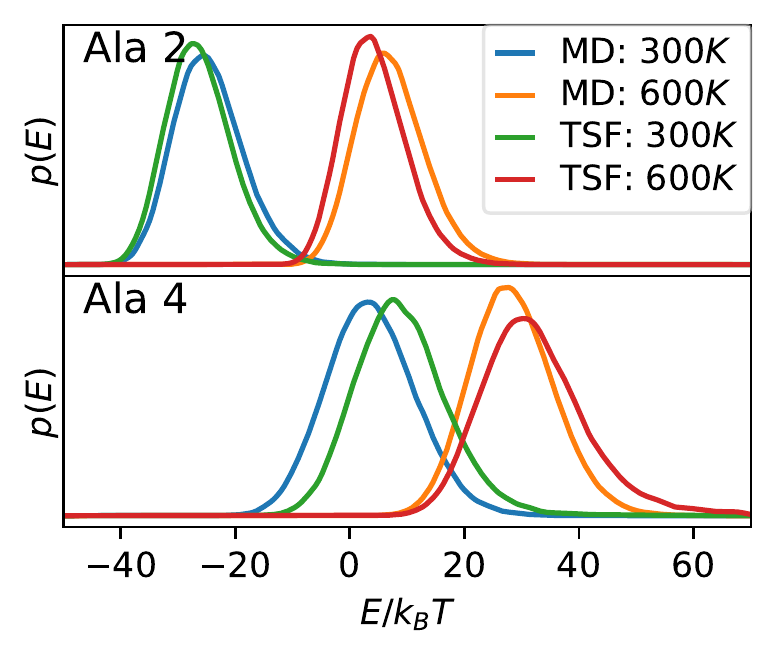}\\
    \includegraphics[width=0.8\columnwidth]{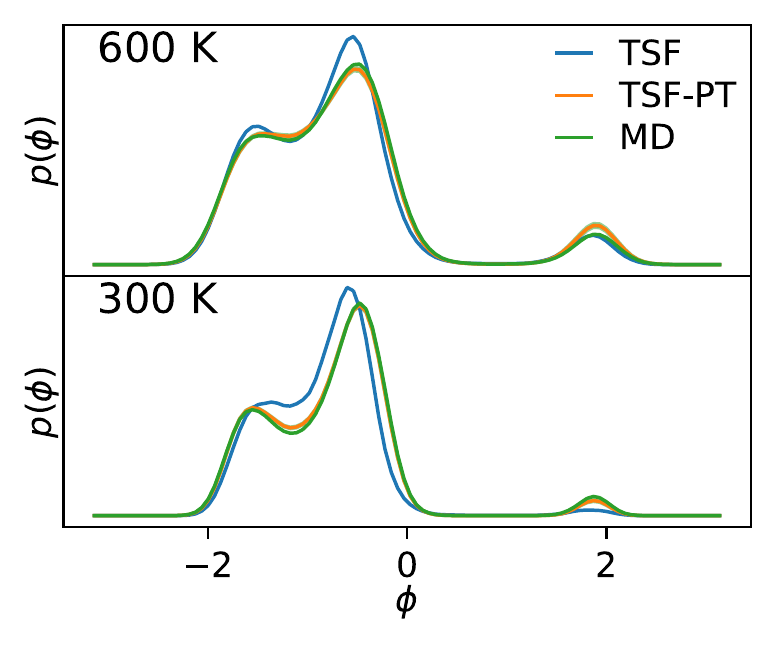}
    \caption{\textbf{Top:} Comparison of histograms of configurational energy between samples generated by the TSF and samples generated by MD. \textbf{Bottom:} Distribution of the slowest timescale in the Ala 2 system generated by different sampling methods involving the TSF.}
    \label{fig:alanine_details}
\end{figure}

Following Liu \citep{liu2008monte}, the effective sample size of
an MCMC sampler can be quantified as $N_{\mathrm{eff}}=N/2\tau$ where
$N$ is the number of samples and $\tau$ denotes the integrated autocorrelation
time $\tau=\frac{1}{2}+\sum_{i=1}^{\infty}\rho_{i}$, with the autocorrelation
function (acf) $\rho_{i}=var(o(\vec{x}^{(0)}),o(\vec{x}^{(i)}))$ of some observable
$o$ that is chosen to be the slowest process in the system. For the
Ala2 system we consider sampling along the $\phi$ angle to be the
slowest process. We define the efficiency of a multi ensemble sampler
(e.g. parallel tempering) simultaneously operating at $M$ copies
of the system as $\eta=N_{\mathrm{eff}}/N=\frac{1}{2M\tau}$ which
quantifies the number of effective (i.e. uncorrelated) samples per
underlying sampling step. We specifically compare REMD to TSF-PT at
$8$ different temperatures in the range of $300\mathrm{\,K}$ to
$600\,\mathrm{K}$ for $10$ independent runs. Trace plots of the
$\phi$ angles at the lowest temperature $300\,\mathrm{K}$, as well
as the acf, are shown in Fig. \ref{fig:Trace-plots-and}. From Fig.
\ref{fig:ala2} we observe, that the system exhibits two metastable
states along this angle. We observe that transitions between these
states happen more frequent in TSF-PT method. Despite being twice
as long, we observe no transitions between the metastable states for
three of the ten independent runs with the REMD method, while all
independent runs transition with TSF-PT. This is also reflected in
the autocorrelation times, which are considerably lower for the TSF-PT
method.

\section{Free energy computation\label{subsec:Free-energy-computation}}
\begin{figure}[b]
\includegraphics[width=1\columnwidth]{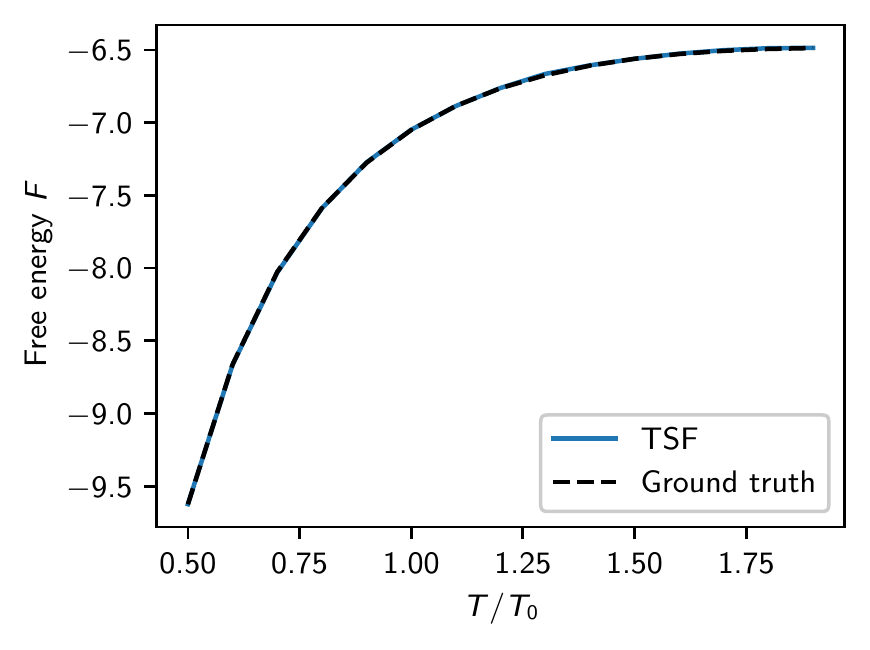}\caption{Free energy dependence on the temperature for a four dimensional system computed with a TSF trained at $T_0$.}
 \label{fig:Free-energy-DW}
\end{figure}

Using the TSF framework, we are able to compute absolute free energies of a system at different temperatures. 
The free energy $F$ is given by
\begin{equation}
F= - \tau \ln(Z_{\tau}),
\end{equation}
where $Z_{\tau}=\int \exp\left(-\frac{U(\vec{x})}{\tau}\right)dx^{3N}$ is the partition function. 
The output distribution $p^{\tau}_X(\vec{x})$ of a normalizing flow is normalized, i.e. $\int p^{\tau}_X(\vec{x}) dx^{3N}=1$. The energy of the output distribution is given by $u_X^{\tau}(\vec{x}) \coloneqq - \log p^{\tau}_X(\vec{x})$ and, hence, $p^{\tau}_X(\vec{x})=\exp(-u^{\tau})$.
We can use samples from the flow to estimate the partition function and therefore the free energy at a given temperature with a TSF.
\begin{align*}
Z_{\tau}&=\int e^{\left(-\frac{U(\vec{x})}{\tau}\right)}dx^{3N}\\
&=\int e^{-\frac{U(\vec{x})}{\tau} + u_X^{\tau}(\vec{x})} e^{- u_X^{\tau}(\vec{x})} dx^{3N}\\
&= \mathbb{E}_{x\sim p^{\tau}_X(\vec{x})} e^{-\left(\frac{U(\vec{x})}{\tau} - u_X^{\tau}(\vec{x})\right)}
\end{align*}
where the last step holds, because the output distribution of the flow is normalized.
As we can sample with the flow and have access to the target energy $U(\vec{x})$ as well as the flow energy $u^{\tau}_D(\vec{x})$, we can compute absolute free energies with the TSF.
We show this for a simple four dimensional test system, where it is easy to compute the total free energy. The target is given by a double well potential and the other dimesnions are given by standard Normal distributions. 
A TSF is trained with the ML-loss objective at temperature $\tau=1$. We see good agreement of the free energy computed uising the TSF with the ground truth (see Fig. \ref{fig:Free-energy-DW}). The ground truth is computed with numerical integration.

\section{Detailed description of the systems and networks\label{subsec:Detailed-description-of}}

\begin{table}[H]
\caption{Setup of alanine dipeptide in implicit solvent}
\begin{tabular}{cc}
\toprule 
\multirow{2}{*}{Force Fields} & \texttt{Amber ff99SB-ILDN}\tabularnewline
 & \texttt{Amber ff99-OBC}\tabularnewline
Number Atoms & 22\tabularnewline
Total simulation time 600K & 10 ns\tabularnewline
Total simulation time 300K & 1 ms\tabularnewline
\multirow{2}{*}{PT temperatures {[}K{]}} & 300.0, 331.2, 365.7, 403.8,\tabularnewline
 & 445.8, 492.2, 543.4, 600.0\tabularnewline
\bottomrule
\end{tabular}
\end{table}
\begin{table}[H]
\caption{Network parameters for the different systems under consideration}
\begin{tabular}{cccc}
\toprule 
System & parameters & coupling blocks & training iterations\tabularnewline
\midrule
Ala2 & 1.9 M & 50 & $3\times10^4$\tabularnewline
Ala4 & 3.3 M & 30 & $3\times10^4$\tabularnewline
Prinz & 30 & 1 & $1 \times 10^4$\tabularnewline
XY-model & 8.5 M & 7 & $1 \times 10^3$\tabularnewline
\bottomrule
\end{tabular}
\end{table}
All experiments were performed on a standard desktop machine equipped
with a \textit{Nvidia GeForce 1080 Ti}, on which one training iteration of the Ala 4 system takes around $0.24$ seconds.
The code will be available at \url{https://github.com/noegroup/bgflow}.

Alanine dipeptide and alanine tetrapeptide were simulated in OpenMM 7.5 \cite{eastman2017openmm} using a Generalized Born implicit solvent model, the Amber ff99SB-ILDN force field \cite{lindorff2010improved}, and a Langevin integrator with 2 fs time step and 1/ps friction coefficient. The time-lagged independent component analysis (TICA) for alanine tetrapeptide was performed using deeptime \cite{hoffmann2022deeptime}. The TICA lag time was chosen as 5 ns based on a subsequent Markov State Model analysis. 

\end{document}